\documentclass[journal,comsoc,onecolumn]{IEEEtran}

\makeatletter
\def\ps@headings{%
\def\@oddhead{\mbox{}\scriptsize\rightmark \hfil \thepage}%
\def\@evenhead{\scriptsize\thepage \hfil \leftmark\mbox{}}%
\def\@oddfoot{}%
\def\@evenfoot{}}
\makeatother
\IEEEoverridecommandlockouts
\usepackage[left=0.7in, right=0.7in, bottom= 1in, top=0.75in]{geometry}
\usepackage{silence}
\usepackage{cite}
\usepackage{amsmath,amssymb,amsfonts}
\usepackage{algorithmic}
\usepackage{graphicx}
\usepackage{textcomp}
\usepackage{xcolor}
\usepackage{booktabs} 
\usepackage{mathptmx} 
\newcommand{\ignore}[1]{}
\usepackage{fancyhdr}
\usepackage[normalem]{ulem}
\usepackage[hyphens]{url}
\usepackage{url}
\usepackage{microtype}
\usepackage{algorithmic}
\usepackage{algorithm}
\usepackage{graphicx}
\usepackage{listings}
\usepackage{algorithm}
\usepackage{mdwmath}
\usepackage{mdwtab}
\usepackage{eqparbox}
\usepackage{mdwmath}
\usepackage{amsmath}
\usepackage{graphicx}
\usepackage{caption}
\usepackage{subfig}
\usepackage{algorithm}
\usepackage{algorithmic}
\usepackage{mathtools}
\usepackage{multirow}
\usepackage[utf8x]{inputenc} 
\usepackage{pbox}
\usepackage{comment}
\usepackage{lscape} 
\usepackage{array}
\usepackage{multicol}
\usepackage{longtable}
\usepackage{mathtools}
\usepackage{makecell}
\usepackage{graphicx}
\usepackage{siunitx}
\usepackage{filecontents}
\usepackage{lipsum}

\def\BibTeX{{\rm B\kern-.05em{\sc i\kern-.025em b}\kern-.08em T\kern-.1667em\lower.7ex\hbox{E}\kern-.125emX}}

\begin{document}

\title{Real time Detection of Spectre and Meltdown Attacks Using Machine Learning}

\author{Bilal Ali Ahmad \\msee17012@itu.edu.pk \\ Information Technology University of Punjab, Pakistan}

\maketitle
\begin{abstract}
Recently discovered Spectre and meltdown attacks affects almost all processors by leaking confidential information to other processes through side channel attacks.These vulnerabilities expose design flaws in the architecture of modern CPUs. To fix these design flaws, it is necessary to make changes in the hardware of modern processors which is a non-trivial task. Software mitigation techniques for these vulnerabilities cause significant performance degradation.In order to mitigate against Spectre
and Meltdown attacks while retaining the performance benefits of modern processors, in this paper, we present a real-time detection mechanism for Spectre and Meltdown attacks by identifying the misuse of speculative execution and side channel attacks. We use hardware performance counters and software events to monitor activity related to speculative execution, branch prediction, and cache interference. We use various machine learning models to analyze these events. These events produce a very distinctive pattern while system is under attack; machine learning models are able to detect Meltdown and Spectre attacks under realistic load conditions with an accuracy of over 99\%. To the best of our knowledge, this is the first-ever run-time detection of Meltdown attack on Intel’s architecture using
hardware/software events and machine learning.
\end{abstract}
\section{Introduction}
\label{int}
The cache-based side channel attacks have evolved since the last decade and imposing significant threats on computer systems security. Researchers are still working on software and hardware mitigation and detection techniques to resolve these issues. On top of that, recently discovered Meltdown and Spectre vulnerabilities also become possible due to these cache-based side channel attacks. All CPUs use speculative execution, out-of-order execution, and branch prediction to enhance performance of CPUs. Spectre\cite{Kocher2018spectre} is a software vulnerability that exploits branch prediction to reveal secret data. The speculative execution, which results from branch mispredictions, leaves observable effects in the cache. Meltdown \cite{Lipp2018meltdown} is a hardware vulnerability which exploits out-of-order execution. It allows any process to read all main memory even when it is not authorized to do so\cite{10.1007/978-3-540-45238-6_6}. The out-of-order memory lookups leave noticeable effects in the cache. Both Meltdown and Spectre attacks influence the cache, which can be observed through CSCAs. These hardware flaws can be fixed only by changing the CPUs hardware [1,2]. These vulnerabilities also impact the way future CPUs will be designed. 

Although, numerous software mitigation techniques have been proposed to mitigate these attack \cite{khasawneh2018safespec}, \cite{wang2018oo7}, \cite{kruger2018vulnerability}, \cite{yan2018invisispec},\cite{kiriansky2018dawg}, \cite{retopline}, \cite{siteisolation},\cite{oleksenko2018you} \cite{spucstore}. These mitigation techniques cause significant performance degradation. In the case of Spectre, it is not possible to fix this issue with a single patch, because Spectre has many variants and each variant required a single patch \cite{news1}. Several software based protection techniques have been proposed to mitigate Spectre attacks. Spectre mitigation techniques have been reported to slow down the performance by 5-12\% \cite{news2}. In the case of Meltdown, software mitigation called ‘KAISER’ was proposed by Maurice et al. \cite{gruss2017kaslr}. It is implemented with the name of KPTI (Kernel Page Table Isolation) in Linux. KAISER has also reported to slow down the CPU. performance\cite{low2018overview}. Also, in \cite{evtyushkin2016jump} Evtyushkin et al. performed branch predictor attack by escaping KASLR. 

In order to make computing devices safe from these hardware flaws, user will need to install patches as soon as they are rolled out. However, it is not possible for all systems to install these patches as soon as they are rolled due to compatibility issues. Despite valiant efforts, mitigation techniques against Meltdown and Spectre attacks are not perfect. Mitigation techniques provide protection against specific variants of Spectre and Meltdown attacks. However, multiple variants of these attacks are coming each day. Also, the effectiveness of these software mitigation techniques comes for the price of performance loss. Therefore, detection techniques can be used as a first line of defense against such attacks and system should apply performance costly mitigation only after detecting these attacks. For detection based prevention strategy to be effective, detection need to be highly accurate, incur minimum system overhead at run-time, and capable of early-stage detection of attacks, i.e., before they complete.

Under these limitations of software mitigation techniques, it is essential to introduce real-time detection technique for Meltdown and Spectre attacks until these attacks are shut down by modifying the hardware of future processors and an all-weather software protection techniques against such attacks do not slow down the performance. We propose real-time detection techniques for Meltdown and Spectre using hardware and software performance counters and machine learning. Our propose detection technique can identify attacking processes with high detection accuracy and minimum system overhead under realistic system load conditions. Similar real-time detection technique has been proposed by Jonas\cite{depoix2018detecting} for the detection of Spectre attack. However, they detected Spectre attacks by identifying cache-based side channel attacks and used only cache related hardware performance counters. Using only cache related hardware events is not reliable indicator for the Spectre attacks detection, Because Spectre attacks also exploit branch predictors. Therefore, we need to include hardware events related to branch prediction to make detection more effective. 

In this paper, we address the problem of accurate and effective detection of Meltdown and Spectre attacks. We present a novel detection technique for attacks which exploit hardware speculation, branch prediction, and out-of-order execution. For Spectre detection, we use hardware performance counters to monitor the caching pattern and branch prediction related events pattern for all running processes. For Meltdown detection, we use hardware performance counters to monitor caching activity for all the running processes and one software event related to page faults. We find that Meltdown attack produces significantly high number of page faults than other benign processes of same computational requirement. We use machine learning models to identify attacking processing by using performance counters collected data. Although thresholding based methods, namely; correlation-based approach, anomaly detection,  can also be used for detection, but smarter adversaries can easily bypass detection techniques that are based on thresholding methods \cite{chiappetta2016real} Followings are the main contributions of this paper:

\begin{enumerate}
    \item We propose a novel detection techniques for Spectre and Meltdown attacks using hardware and software performance counters.     
    \item We find effective hardware performance counters for the detection of Spectre attack. We demonstrate the effectiveness of our detection technique under realistic system load conditions, i.e., under No load, Average load and Full load conditions, with high-performance accuracy and minimal performance overhead. 
    \item We propose, to the best of our knowledge, a first-ever detection technique for Meltdown attack using hardware/software performance counters and machine learning. We demonstrate that the out-of-order memory lookups in case of Meltdown attack generate significantly high number of page faults which can be used as a better indicator for Meltdown attack detection along with cache related hardware events. 
    \item We find hardware and software related events that are directly related to attacks, exploit hardware speculation, branch prediction, and out-of-order execution. Therefore, our proposed detection technique can also be used to detect attacks which exploits performance enhancement features of CPUs. 
    \item We provide experimental results and discussion on detection accuracy and system-wide performance overhead for selected machine learning models. We show that the Machine learning models are successfully able to identify processes which use Meltdown and Spectre attacks. We also show that computationally less expensive linear machine learning models provide same detection accuracy as of computationally expensive ML models like CNN which is based on deep learning (Convolutional Neural Network). 
    \item We demonstrate the circumstances to select best performing machine learning models and constraints related to the utilization of HPCs.
\end{enumerate}

In section \ref{background}, we give the necessary background to understand Meltdown and Spectre attacks such as cache-based side channel attacks. In section \ref{csca} , hardware speculation attacks, in section \ref{fraes}, out-of-order execution, in section \ref{ffaes}, Meltdown attack in section \ref{metldown}, Spectre attack in section \ref{spectre} and hardware/Software performance counters attacks in section \ref{hpc}. Section \ref{bck} presents related work. Section \ref{dectetionscheme} presents the working principle of run-time detection mechanism. Section \ref{sec:experiments} provides experimental evaluation and discussion. Section \ref{conclusion} concludes this paper.


\section{Background Knowledge}
\label{background}

This section provides the fundamental background needed to understand the methodology, experiments and results of our proposed solution.

\subsection{Cache-based side channel attacks}
\label{csca}
Side channel attacks are the attacks which observe the side effects generated by optimization methods used in computer systems. When the state of cache is shared between two different programs for execution, there might be a potential risk of timing channel as the victim and attacker might be sharing the same resources and attacker can observe the victim’s execution with minute timing variations because the timing of one program is dependent on the execution of the other program \cite{hu1992reducing}, \cite{tsunoo2002crypt}. Cache-based side channel attacks are also known as timing attacks. These attacks work by exploiting the cache timing information. Loading something from the main memory takes a lot more time than loading something from the cache. Attackers observe how long it takes to access the data from a specific memory address. From this timing information, the attacker can conclude that either accessed data is present in cache or not. Many cache-based side channel attacks such as PRIME+PROBE\cite{liu2015last}, EVICT+TIME\cite{10.1007/11605805_1}, FLUSH+RELOAD\cite{184415}, and FLUSH+FLUSH\cite{Gruss:2016:FFS:2976956.2976976} have been proposed to leak secret information from cryptographic. 

CSCAs also plays a role in making Spectre and Meltdown attack possible. Both these attacks are not possible without these side channel attacks. We can use any cache-based side channel attacks to perform Spectre and Meltdown attacks, but the original work of Meltdown and Spectre used FLUSH+REOLAD attack. Flush+Reload \cite{184415} falls under the classification of trace-driven attacks because it relies on presence of page sharing. To add to the problem of inclusive caches, x-86 architecture provides privileged instructions, such as clflush instruction, for flushing the memory lines from all cache levels, including the last level cache (LLC), which proves to be a major threat and core advantage for attacks using Flush+Reload technique. In the first phase of flushing, the attacker flushes (evicts a shared cache line) using clflush instruction. After flushing the cache line, attacker remains in the idle state and lets the victim operate. In the step of Reload, it observes the timing information by reloading the shared cache line. The timing information reveals the interest of victim program. Stealth reload indicates that this cache line was affected by victim and slow reload shows that the victim did not access it. This mechanism is exploited by attacker to leak secret information.

\subsection{Hardware speculation}
\label{fraes}
Hardware speculation is a technique used to make instructions execution rate high through dynamic branch predictions and dynamic scheduling of instructions. Branch predictors try to guess which way the branch will go before this is actually known by CPU. Branch predictors uses the history of taken or not taken branches to predict the outcome of conditional branches at run time. Branch predictors are used to enhance the performance of conditional and jump branch instructions. Branch predictor guess the outcome of branch instructions. Usually, it speculates the control flow of branch instructions using history of the same branch instructions. In case of conditional branch instructions, usually two way branching is used for conditional branch instructions. The outcome of conditional branch can either be taken or not taken. It is not known either branch will be taken or not taken until the condition branch has been evaluated. Without branch prediction, processor has to stall in pipelined stage until the conditional branch has been calculated that will be a wastage of processor cycles and no other instruction can enter in fetch stage of pipeline. Therefore, branch predictors are used to avoid this waste of processor time. During program execution, branch predictor guess either conditional jump is likely to be taken or not. The branch which is predicted to be most likely, is fetched and speculatively executed before the result conditional is actually known. If in later stage, it is known that the guess of predictor was wrong, processor discards the result of speculatively executed instructions\cite{gruss2017kaslr} and starts to execute correct branch. If branch prediction was correct, it will significantly increase the instructions execution rate and hence performance. 

All existing processors use branch predictors to guess the outcome of direct branches, indirect jumps and calls \cite{news4}. All variants of Spectre attack exploit direct branches, indirect branches and calls. Several processor components are used for predicting the outcome of branches. The Branch Target Buffer (BTB) keeps a mapping from addresses of recently executed branch instructions to destination addresses \cite{Lipp2018meltdown}. Spectre attack mistrains the history of BTB to execute unauthorized instructions speculatively.

\subsection{Out-of-order execution}
\label{ffaes}

Out-of-order execution is an optimization method used by modern processors to achieve maximum utilization of execution units available in a CPU. In Out-of-order execution, instructions are fetched in compiler generated sequence. But instructions can be executed in order or Out-of-order depending on the data hazards and structural hazards between instructions. Instructions can execute Out-of-order, but they complete in order only \cite{abu2019spectre}. A processor having out-of-order execution functionality does not wait for the instructions to complete their execution in sequential order. Preceding instructions start executing if all necessary operands and functional units are available without waiting for the previous instructions to complete their execution. Meltdown attack exploit this feature of modern processors by using out-of-order memory lookups. It is explained in more details in section \ref{metldown}.

\subsection{Meltdown Attack}
\label{metldown}
Meltdown is a hardware vulnerability which affects Intel x86 microprocessors, IBM POWER processors, and some ARM-based microprocessors. Meltdown \cite{Lipp2018meltdown} exploits out-of-order execution of microprocessors to leak secret information from user space and Kernel space of same process and other processes. It bypasses page level permission check through out-of-order execution of unprivileged instructions. Memory protection mechanism is provided in all processors through operating system that prevents a user level program to read data from kernel space or from any other use level program. However, side effects generated by out-of-order execution makes it possible to bypass this Memory protection mechanism. Even an unprivileged user level program can read all main memory with Meltdown attack. It is a two-step attack. In first step, meltdown bypass the memory isolation by executing unprivileged instruction out-of-order. In second step, it performs cache-based side channel attack to observe footprints of accessed data from cache. Listing 1 shows the code snippet used to first raise an exception or segmentation fault and in line 3 accessed the probe array. 

\begin{lstlisting}[caption=Meltdown Attack]
1.raise_exception(); 
2.//the line below is never reached
3.access(probe_array[data * 4096]);
\end{lstlisting}

When exception occurs, control flow of program shifts to operating system and next instructions do not get executed. In above example, array will not be accessed because control flow moves to operating system and terminate the process. But due to out-of-order execution, line (3) may be executed out-of-order before kernel terminates the application. But the instructions executed out-of-order will not be committed due to exception. The instruction which will be executed out-of-order do not have any effect on registers and memory. However, they have microarchitectural side effects which can be observed through caches. During out-of-order memory lookups, referenced memory is stored both in registers and cache. If out-of-order executed instruction does not commit, the results of instruction are discarded from memory and registers. But the contents of the referenced memory do not remove from cache until the cache replacement policy applies. Once the data are in cache, Meltdown attack use a cache-based side channel attack such as FLUSH+RELOAD to determine either a specific memory location is cached or not \cite{184415}. Based on this observation, attacker can leak the secret information by performing Flush+Reload continuously. Meltdown attack use exception handing to control the flow of code from being transferred to operating system, which saves the application termination. However, we can still monitor this exception signal through software events of Perf \cite{news5}. This exception or segmentation fault signal is a good indicator for the detection of Meltdown attack. We discuss more about this software performance counter in \ref{sec:experiments}.   

 \subsection{Spectre Attack}
 \label{spectre}
 Unlike meltdown attack, Spectre \cite{Kocher2018spectre} does not generate any exception or segmentation fault instead it exploits branch prediction to bypass isolation between user level processes. Spectre has multiple variants. But, we discuss Spectre variant 1 and variant 2 that are used in original work of Spectre attack. Other variants of Spectre such as BranchScope\cite{evtyushkin2018branchscope},Netspectr\cite{schwarz2018netspectre} Speculative buffer overflows\cite{kiriansky2018speculative}, ret2spec\cite{maisuradze2018ret2spec},ExSpectre\cite{wamplerexspectre} are also related to Spectre variant 1 and Spectre variant 2. Spectre attacks mistrain the branch prediction units of almost all processors. Branch predictor are use for prediction of conditional branch instruction, indirect branch instructions and return stack buffer. Spectre variants are available for all three types of branch instructions. For condition branches, branch predictor predicts whether a conditional branch, such as if else instructions, will be taken or not taken. Branch predictor guess the direction of conditional branch depending on the history of branches. Similarly, branch predictor make guess for indirect branches and calls \cite{abu2019spectre}. All these branch instructions are exploited by Spectre attack to leak secret information.
 
Spectre attack is also a two-step attack. In step one, attacker mistrains the branch predictor of CPU to speculatively execute unprivileged instructions. In second step, perform cache-based side channel attack to leak information unauthorized reference memory. Listing 2 shows the code snippet of Spectre variant 1. 

\begin{lstlisting} [caption= Spectre variant 1 code snippet ]
1.if (x < array1_size) 
2.  y = array2[array1[x] * 4096];
\end{lstlisting}

In variant 1 of Spectre attacks, the attacker mistrains the branch predictor unit of CPU’s to mispredicting the direction of conditional branches. Attacker mistrains the CPU’s branch predictor to execute unprivileged branch instructions which were not executed otherwise. The attacker first trains the branch predictor to always take decision of conditional branch as a taken branch and after training the branch predictor, attacker executes out of bound instruction. In the example above, assume that the variable x contains attacker-controlled data. To ensure the validity of the memory access to array1, the above code contains an if statement whose purpose is to verify that the value of x is within a legal range. We show how an attacker can bypass this if statement, thereby reading potentially secret data from the process’s address space. 

First, during an initial mistraining phase, the attacker invokes the above code with valid inputs, thereby training the branch predictor to expect that the if will be true. Next, during the exploit phase, the attacker invokes the code with a value of x outside the bounds of array1. Rather than waiting for determination of the branch result, the CPU guesses that the bounds check will be true and already speculatively executes instructions that evaluate array2[array1[x]*4096] using the malicious x. Note that the read from array2 loads data into the cache at an address that is dependent on array1[x] using the malicious x, scaled so that accesses go to different cache lines and to avoid hardware prefetching effects. When the result of the bounds check is eventually determined, the CPU discovers its error and reverts any changes made to its nominal micro-architectural state. However, changes made to the cache state are not reverted, so the attacker can analyze the cache contents and find the value of the potentially secret byte retrieved in the out-of-bounds read from the victim’s memory.

In Spectre variant 2, attacker exploits Branch Target Buffer to mispredict the result of indirect branch instructions. Attacker mistrains the Branch Target Buffer. To mistrain the BTB, the attacker finds the virtual address
of the data in the victim’s address space, then performs
indirect branches to this address. This training is done from the attacker’s address space. After training BTB, attacker mispredict a branch from an indirect branch instruction to the victim's address space that results in speculative execution of instruction from victim's address space. As before, while the effects of incorrect speculative execution on the CPU’s nominal state are eventually reverted, their effects on the cache are not, thereby allowing the speculatively executed instruction to leak sensitive information via a cache side channel.

Spectre variant 1 exploits Branch-direction predictor and Spectre variant 2 use Branch-target buffer. Although both variants of Spectre mistrain different branch predictor units, but both variants perform carefully crafted branch mispredictions after every training phase.Spectre attacks generate significantly large branch mispredictions as compared to total number of branch instructions in program. Therefore, these branch mispredictions are good indicator for the detection of Spectre attacks along with cache based side channel attack. we explain this through experimental results in section \ref{sec:experiments}.

 \subsection{Hardware Performance Counters (HPCs)}
 \label{hpc}
 Hardware performance counters have been available in all microprocessors for more than a decade. Hardware performance counters are special purpose registers built in the microarchitecture of CPUs \cite{news5}. HPCs are available as a hardware registers that monitor data about processor events. These hardware performance counters are used to monitor CPU activity during program execution like total cycles, retired instructions, branch prediction, cache missed and cache hits. Modern Processors support hundreds of hardware performance counters, but physically only few hardware counters are available  Hardware performance counters are mainly used for debugging and performance analysis of programs under execution.However, researchers are also using HPCs for security applications e.g. malware detection, cache-based side channel attacks detection. Performance counters has been used in research work \cite{wang2016reusing},\cite{leng2017hardware}, \cite{rasoolzadeh2014estimating} for dynamic profiling
and intrusion detection
 
 \subsection{Software performance counters}
Almost all operating systems support software performance counters, such as, page faults, major page faults, minor page faults and invalid page faults. Unlike HPCs, software events are not a microarchitecture specific events, but specific to operating systems. All operating systems provides tools to monitor software performance counters. Like, Linux provides a performance monitoring tool known as perf \cite{news5} to monitor both hardware and software events. Unlike PAPI, Perf supports both hardware and software performance counters \cite{de2010new}. Perf have been used in research \cite{sturges2009using} for performance analysis of applications. In this research paper, we have used PAPI for the detection of Spectre attacks and Perf tool for the detection of Meltdown attacks. We will discuss further about selected hardware and software events in section 4.2.  

\section{Related Work}
\label{bck}
This section summarizes the state-of-the-art on cache-based side channel attacks detection techniques (using machine learning and hardware performance counters) and explains their issues and limitations. Since the evolution of cache-based side channel attacks, researchers have proposed various detection techniques. Most of them have limitation and issues for practical implementation. Only some proposed detection techniques provided an evaluation of their proposed methodologies under realistic system load conditions. In \cite{chiappetta2016real} Chiappetta et al. used Neural Networks and HPCs to detect FLUSH+RELOAD attack on RSA, AES, and ECDSA. They used PAPI to monitor HPCs events related to L3 cache regularly \cite{news3}. They used this monitored data for detecting FLUSH+RELOAD attack using Neural Networks. Although, they did not provide experimental evaluation of detection technique under realistic system load conditions and also did not comment on performance overhead of their detection technique. Neural Networks acquire high performance cost during training and detection stage. Therefore, they are not appropriate for detection of side channel attacks. In \cite{Mushtaq:2018:NCS:3214292.3214293} Mushtaq et al. used linear machine learning models and HPCs to detect FLUSH+RELOAD and FLUSH+FLUSH attacks on RSA and AES respectively. They provided experimental evaluation of detection techniques using realistic system load conditions and analyze results on detection accuracy, speed and system-wide performance overhead for linear machine learning models. Their results show that linear ML models incur low performance overhead and are suitable for run time detection of side channel attacks. In \cite{mushtaq:hal-01879950} Mushtaq et al. used machine learning and hardware performance counters to detect Prime+Probe attack on AES implementation. They have shown good experimental results under realistic system load conditions. In recent years, similar research work such as \cite{bazm2018cache}, \cite{alam2018rapper}, \cite{zhang2016cloudradar}, \cite{inci2016co} has been done to detect cache-based side channels attacks and malware detection using HPCs and machine learning models. These research works shown the potential HPCs have for detecting cache-based side-channel attacks.

Meltdown and Spectre are recently discovered attacks. Only few detection techniques are proposed to detect Spectre attacks and to the best of our knowledge, no detection technique is reported for Meltdown detection. In \cite{depoix2018detecting} Jonas et al. proposed a real-time detection technique for Spectre attacks using hardware performance counters and Neural Networks. They built detection model by training Neural Networks using HPCs data on benign and Spectre processes. They detected the Spectre attack by identifying the cache-based side channel attack which is similar to the work of Chiappetta \cite{chiappetta2016real} and Mushtaq \cite{Mushtaq:2018:NCS:3214292.3214293}. Their proposed detection technique has multiple issues and limitations. Firstly, they detected Spectre attacks by identifying Cache side channel attacks without considering the main root cause of Spectre attacks. They used hardware events such as L3 cache misses, L3 cache accesses, the total number of instructions and total CPU cycles which are related cache-based side channel attacks only. Therefore, their detection technique is only for cache-based side channel attacks detection. In \cite{yao2019towards} Yao et al. showed that cache related hardware events are not a good indicator for cache timing attacks detection and smarter adversaries can easily spoofed cache misses and accesses patterns to bypass existing detection techniques. Therefore, detection Spectre only by identifying side channel attack is not a strong detection technique. Secondly, their detection technique is performance costly due to Neural Networks high resources requirement during training and detection phase. In\cite{mushtaq:hal-01876792} Mushtaq et al. provided experimental evaluation and comparative analysis on the use of various Machine Learning (ML) models for detecting Cache-based Side Channel Attacks (CSCAs) in Intel’s x86 architecture. They showed that linear machine learning models provide higher detection accuracy under realistic system load condition with minimum performance overhead. Therefore, it is not suitable to use Neural Networks when linear ML models can provide excellent results with minimum performance overhead. 
 
As we discussed in section \ref{spectre}. The Spectre is a two steps attack. In the first step, the attacker exploits the branch predictions to execute unprivileged instruction speculatively, and in the second phase, the attacker executes side channel attack to extract secret information. In order to detect all variants of Spectre attacks, we need to consider the main root cause of these attacks. The main root cause of Spectre attacks is mis training of branch predictors units. Therefore, we also need to add hardware events related to the first step of the Spectre attack. In \cite{vougioukas2019brb} Vougioukas et al. showed that branch predictors increase their misprediction rate by as much as 90\% on average when used by the attacks which exploit branch prediction. The Spectre attack also exploits branch prediction. Therefore, the total branch instructions and total branch miss-prediction can prove to be a good indicator for the detection of the Spectre attacks. The novel contribution of our research work is that we are using hardware events which are directly related to the Spectre attack along with cache-based side channel attack. In case of Meltdown attack, to the best of our knowledge, we are the first one to propose the detection technique. Although in \cite{depoix2018detecting} Jonas et al. mentioned that we can detect meltdown attack by identifying side channel attack. However, they did not provide the experimental evaluation of meltdown detection technique. They also did not talk about the primary cause of Meltdown Vulnerability. Meltdown attack produces significantly large page faults than other benign processes of the same computational requirement. Page faults can be a good indicator for the detection along with cache-related events. 

 \section{Proposed Run-time Detection Mechanism}
 \label{dectetionscheme}

\begin{figure*}[htp]
\label{blockdiagram}
\centering
\includegraphics[width=0.80\textwidth]{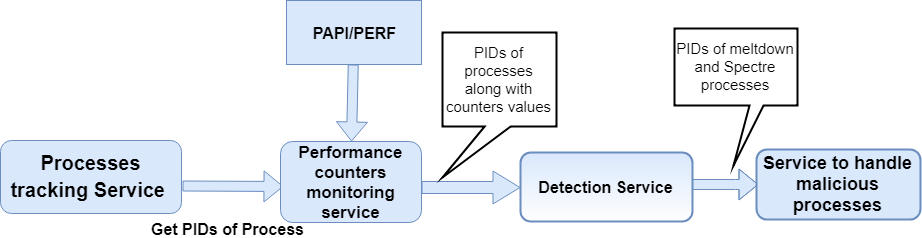}
\caption{Abstract view of detection mechanism.}
\label{fig:detection}
\vspace{-2ex}
\label{fig:blockdiagram}
\end{figure*}
\vspace{2ex}

\subsection{Detection Methodology}
\label{sec:method}

Figure \ref{fig:blockdiagram} shows the methodology of our proposed detection technique for Spectre and Meltdown. Our proposed detection technique tracks all the processes and monitor counters values for each process and pass counters values to detection technique. Detection Service sends the PIDs of malicious processes to an application which decides about these malicious processes such as Meltdown and Spectre.As we propose only detection technique and does not propose mitigation, therefore mitigation technique can decide with to do with these malicious processes. 

To monitor malicious processes, the process tracking service tracks the processes as soon as they start or stop. It also gets the PIDs of the concurrent processes. Processing tracking service sends this information to performance counter monitoring service. Performance counter monitoring service perform run time profiling of these processes by monitoring performance counters values. Performance counters monitoring service sends the processes PIDs along with their monitored counter values to Detection service. Detection Service decides whether the crossponding process is malicious or benign. Detection Service forwards the PIDs of malicious processes to the application which decides that what to do with these malicious processes. We provide more information on detection service in section \ref{rtd}. 

\subsection{Detection Service}
\label{rtd}
This section summarizes details on our propose detection technique for Meltdown and Spectre attacks. Machine learning has already been used for security applications such as in \cite{chiappetta2016real},\cite{Payer:2016:HPD:2990910.2990919}, \cite{mushtaq:cel-01824512}, \cite{10.1007/978-3-319-66939-7_17}, \cite{cryptoeprint:2017:564}, \cite{8588177},\cite{buczak2016survey}. Machine learning models are used to find patterns in data by identifying input output relationship.ML models are used for binary and multi classification of data \cite{latexcompanion}. We use ML models for binary classification problem because our detection technique classifies processes either as malicious or benign. ML models can be divided into types of linear and non-linear models. We select machine learning models which incur low performance overhead and offer easy implementation feasibility. We select three linear machine linear models and we also select Neural Networks for experimental evaluation because we compare its results with other linear ML models. We perform experiment with 3 linear models and 1 non-linear model namely; LR, SVM, LDA and CNN, due to following reasons. Although, it is the most important one, classification accuracy is not the only parameter to consider while deploying a ML model for detection mechanism. We reason that the other most important parameter to examine while comparing ML models is their implementation feasibility. ML models should also be able to quickly provide their decision on detection based on the profiling of processes using HPCs. Moreover, the performance overhead caused by ML models for detection should be reasonable. We retrieve the features using performance counters and train these models on them. Counters provides real time information of all concurrent processes.

Our detection mechanism consists of three phases namely; the Training phase, Run-time profiling and Classification phase and detection phase.

\subsubsection{Phase-I: Training of ML Models}
In training phase, we train the machine models being used for detection technique of Spectre and Meltdown. We profile Meltdown/Spectre and other benign processes under Attack and No Attack scenarios and variable load conditions. We collect performance counter values at sampling rate of 100ms. We monitor suitable hardware/software counters (listed in section \ref{selhpc}) for profiling, which are directly related for differentiating benign processes from Meltdown and Spectre processes, separately. We train the machine learning models with a data set of 100,000 samples. We mixed the data sample from benign and attacker processes and train ML models on these labeled data. We perform cross validation using K−fold cross validation technique

\subsubsection{Phase-II: Run-time Profiling}
 In the second phase, we monitor hardware and software events at run time. Sampling frequency is an important factor for detection technique, as it has a direct impact on performance overhead of detection. We select the sampling time period of 100ms to incur minimum detection overhead. However, greater sampling frequency provides high detection speed, but it comes for the price of performance overhead. Therefore, we select a coarse grain profiling mode, it takes samples at relatively low frequency, which takes a little longer to detect, but incurs much less performance cost. 
 
\subsubsection{Phase-III: Classification \& Detection.}
 In this phase, we pass the data collected in last phase to trained ML models in real time. Based on this data, train ML models classify processes either as benign or malicious. We provide details of detection accuracy, False positives and False Negatives for each ML model in section \ref{sec:experiments}.


 \vspace{-1ex}
 \subsection{System Model}
 \label{sysmod}
We demonstrate the effectiveness of proposed detection mechanism on Intel’s core i3 − 2120 CPU running on Linux Ubuntu 16.04.1 with kernel 4.13.0−37 at 3.30-GHz. Our thread model consists of detecting attacks which exploits hardware speculation, branch predictors, out-of-order execution and cache-based side channel attacks in Intel’s x86 architecture. There are many high-level software libraries/ APIs that are used to monitor hardware performance counters and software events HPCs such as; PerfMon, OProfile, Perf tool, Intel Vtune Analyzer, and PAPI. We use PAPI and Perf to monitor performance events on intel’s core I3 machine. We use PAPI to extract events related to Spectre attacks and Perf to extract events related to Meltdown attack. These performance counters are used to train machine learning models. To train machine learning models, we produced data set from various benign processes and meltdown/Spectre processes using these hardware and software performance counters. We monitored performance counters of each process and labelled them as benign or malicious. We provided more information about it in section 4(D).

 \section{Selected Hardware Events as Useful Features}
 \label{selhpc}

In this section, We provide overview of selected hardware and software performance counters used for run time detection of Meltdown and Spectre attacks. Our proposed detection technique identifies the data patterns of these vulnerabilities that are not similar to expected benign processes (which require same computational requirement). Our detection technique learns the expected behavior of system under normal system conditions and under these attacks by using machine learning and performance counters as useful features. Our detection technique uses hardware/software performance counters and machine learning models to observe patterns generated by these attacks and identify processes which use Spectre and Meltdown attacks. Since detection mechanisms can only approximate the system behavior, they can be inaccurate and lead to false positives or false negatives at run-time. Moreover, they can also slowdown program execution due to detection overhead. We use all these parameters as evaluation metrics for the proposed detection technique. We first describe our system mode.
 
\begin{table}[httb]
\renewcommand{\thetable}{\arabic{table}}  
\vspace{1ex}
  \centering
    \caption{Selected Performance counters for Spectre}
    \begin{tabular}{|c|c|c|}
    \hline
    \textbf{Scope of event}  & \textbf{Hardware event} &  \textbf{Feature ID} \\  \hline
      L3 cache&	Total cache misses & L3\_TCM \\ \hline
      L3 cache&	Total cache access & L3\_TCA \\ \hline
      System wide&	Total branch instruction & BR\_INS  \\ \hline
      System wide& 	Branch Miss-Predictions & BR\_MSP  \\ \hline
      System wide&	Total number of instructions & TOT\_INS \\ \hline
  \end{tabular}
 \label{tableSpectre}
\end{table}

The PAPI and Perf API provide 100+ performance counters for intel’s core i3 processors. We use these API’s to monitor system wide, per-process and per CPU events. Our proposed detection techniques classify a process either as benign or malicious from all running processes. Therefore, we use PAPI and Perf to monitor per process events. We attach these APIs to each running process and extract performance counters values specific to each process. our detection method uses these performance counters data as features to ML models. In order to select relevant performance counters for Spectre and Meltdown, we performed experiment using various hardware and software events. We select system wide profile for benign and malicious processes such as Meltdown and Spectre. To train ML model for Spectre detection, we generate data set using four processes. One process is an implementation of Spectre and other 3 processes are benign processes. Similarly, for Meltdown detection, we produce training data set using one Meltdown process and other three processes having same computational requirement. We select these events based on three important factors: 1) Their relevance to Meltdown and Spectre attacks 2) Their potential to provide better classification 3) Only select minimum possible counters to make detection overhead minimum. We provide more details on selected performance counters for Meltdown and Spectre in section \ref{spectredetction} and \ref{Meltdowndetction} respectively. 

\begin{figure}[htbp]
\centering
\includegraphics[width=0.4\textwidth]{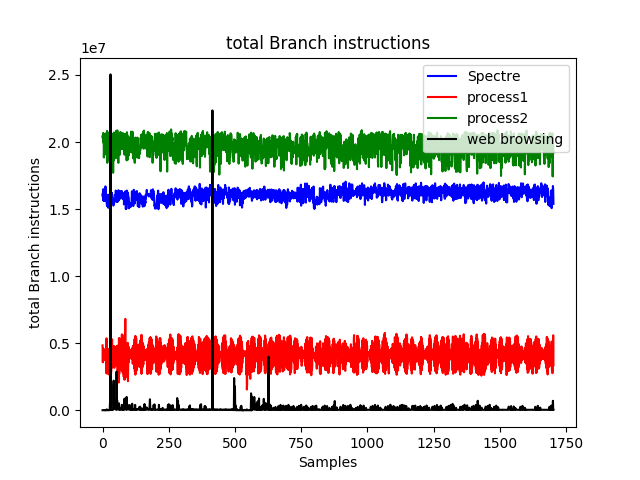}
\caption{Total branch instructions}
\label{fig1}
\vspace{-2ex}
\end{figure}
\begin{figure}[htbp]
\centering
\includegraphics[width=0.40\textwidth]{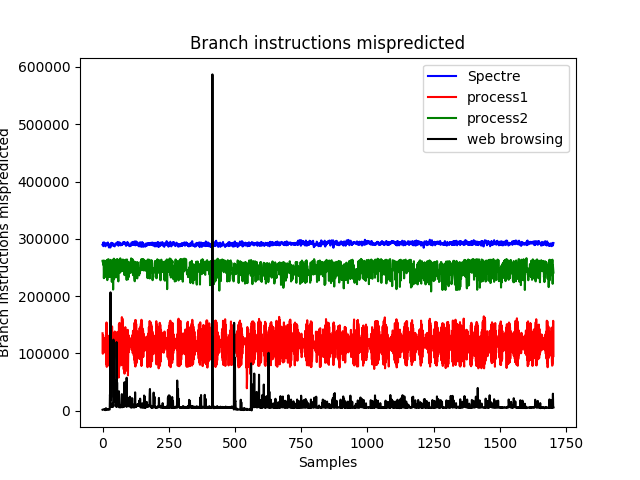}
\caption{Total branch instructions mispredicted}
\label{fig2}
\vspace{-2ex}
\end{figure}

\subsection{Selected Performance Counter for Spectre}
\label{spectredetction}

As mentioned in section \ref{spectre}, attackers perform Spectre in two phases. In first phase, attacker mistrains branch instructions to execute unprivileged instruction speculatively and in second phase, attacker execute cache-based side channel attack to leak secret information. Based on the features selection criteria, discussed in section \ref{selhpc}, we select performance counters related to both phases of Spectre attacks. Main root cause of all variants Spectre attacks are mis training of branch instructions. Spectre variant 1 exploits Branch-direction predictor and Spectre variant 2 use Branch-target buffer. Although both variants of Spectre mis train different branch predictor units, but both variants perform carefully crafted branch mispredictions after every training phase. Therefore, we select two hardware events related to the first phase of Spectre attacks such as total branch instructions and total branch mispredictions. Also, in \cite{vougioukas2019brb} Vougioukas et al. showed that branch predictors increase their misprediction rate by as much as 90\% on average when used by the attacks which exploit branch prediction like Spectre attacks.  

\begin{figure}[htbp]
\centering
\includegraphics[width=0.40\textwidth]{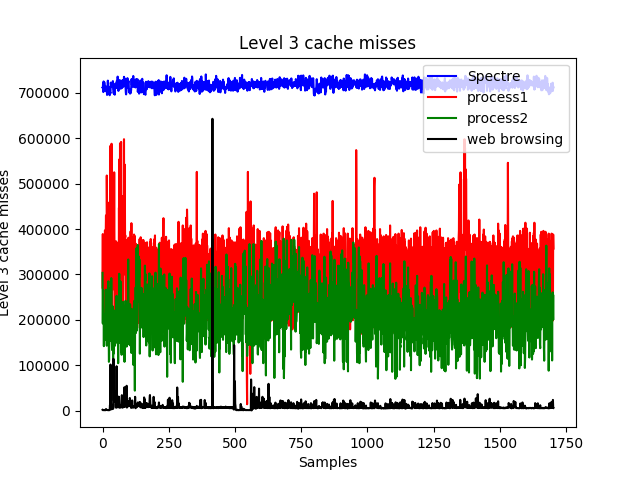}
\caption{Level 3 cache accesses}
\label{fig3}
\vspace{-2ex}
\end{figure}

Figure \ref{fig2} shows magnitude of total branch instructions for three benign processes and one Spectre processes. Figure \ref{fig3} shows the branch mispredictions generated by all processes. Spectre attack produces significantly large branch mispredictions as compared to total number of branch instructions of Spectre attack. Also, it generates comparatively higher branch mispredictions than other benign processes. Result of Vougioukas et al. and our experimental results shown in Figure \ref{fig2} and Figure \ref{fig3} prove our intuition for selection of branch related hardware events as good features for ML models.

\begin{figure}[htbp]
\centering
\includegraphics[width=0.40\textwidth]{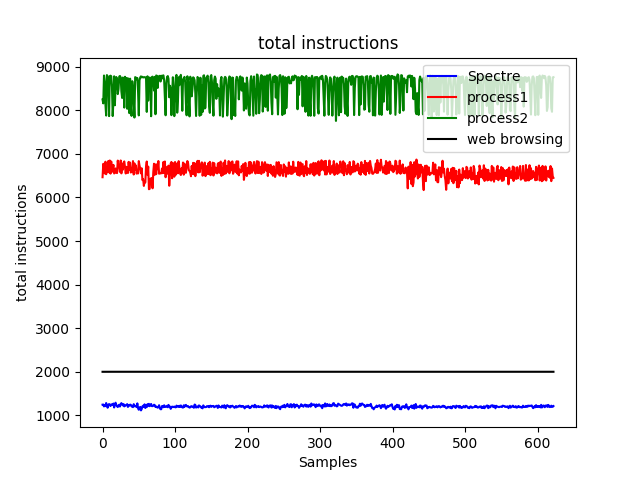}
\caption{Level 3 cache misses}
\label{fig4}
\vspace{-2ex}
\end{figure}
\begin{figure}[htbp]
\centering
\includegraphics[width=0.40\textwidth]{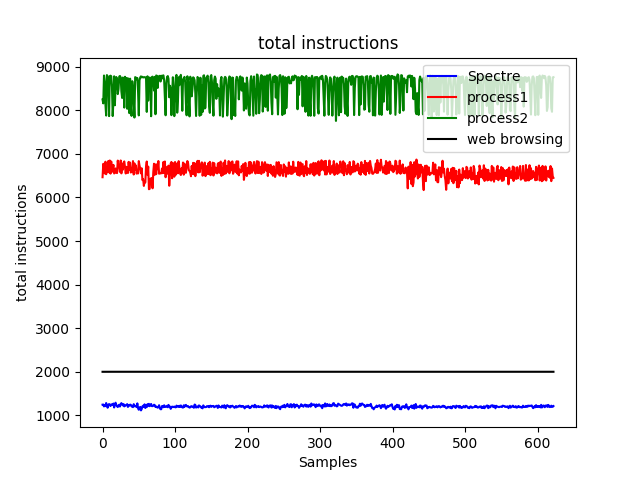}
\caption{Total number of instructions}
\label{fig5}
\vspace{-2ex}
\end{figure}

We also select hardware events related to second phase of Spectre attacks to strengthen detection technique and also to minimize False positives and False negatives. In second phase, Spectre attacks performs cache-based side channel attack, like FLUSH+RELOAD, to extract confidential information from caches. As mentioned in section \ref{csca}, to execute FLUSH+RELOAD, attacker continuously flush the cache lines and checks after some time if it has been accessed by victim since the last flush. By constantly performing cache flushing and reloading, attacker executes a lot of cache accesses and of which a lot will be cache misses, in a repetitive pattern. Attacker access the caches in a malicious way. Therefore, attacker generates a significantly higher cache miss rates while performing Flush+Reload. PAPI supports Hardware performance counters related to cache misses and cache accesses for all levels of cache. We select L3 cache misses and cache accesses because flushing cache line from L1 cache also removes the content from all levels of cache. Figure \ref{fig4} and \ref{fig5} show the total number of cache misses and total cache access generated by benign and Spectre process. As depicted from graphs, Spectre attack produce significantly larger L3 cache misses and L3 cache accesses as compared to other benign processes. These experimental results show that cache misses and cache accesses are good indicators for detecting Spectre attack. 

\begin{figure}[htbp]
\centering
\includegraphics[width=0.40\textwidth]{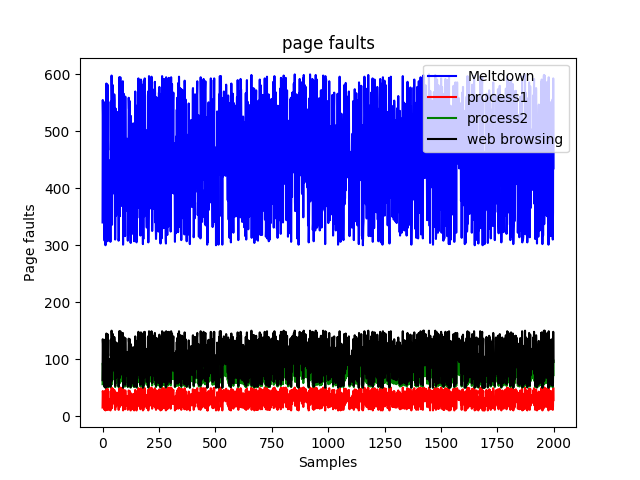}
\caption{Total number of page faults}
\label{fig6}
\vspace{-2ex}
\end{figure}

In addition to L3 cache misses, cache access and branch related events, we also select total number of instruction event because it shows the workload a specific process put on the CPU to generate related cache misses, cache accesses and branch miss predictions. Because Spectre attack consists of a shorter loop which constantly performs branch mispredictions and cache accesses. The rate of cache misses and branch mispredictions in relation to the total number of executed instructions is likely to be higher for Spectre attacks as compared to other benign processes. Figure \ref{fig6} shows that the Spectre process puts a very small workload, in the form of total number of executed instructions, on CPU but still generates higher cache misses, cache accesses and branch miss-predictions. Relevant hardware events selected for Spectre attacks are listed in Table \ref{tableSpectre}. 
 
\subsection{Selected Performance Counter for Meltdown }
\label{Meltdowndetction}

\begin{table}[htbp]
\renewcommand{\thetable}{\arabic{table}}  
\vspace{1ex}
  \centering
    \caption{Selected Performance counters for Meltdown}
    \begin{tabular}{|c|c|c|}
    \hline
    \textbf{Scope of event}  & \textbf{Hardware event} &  \textbf{Feature ID} \\  \hline
      L3 cache&	Total cache misses & L3\_TCM \\ \hline
      L3 cache&	Total cache access & L3\_TCA \\ \hline
      System wide& 	Total page faults & page\_faults  \\ \hline
      System wide&	Total number of instructions & TOT\_INS \\ \hline
  \end{tabular}
 \label{tablemltdown}
\end{table}

\begin{figure}[htbp]
\centering
\includegraphics[width=0.40\textwidth]{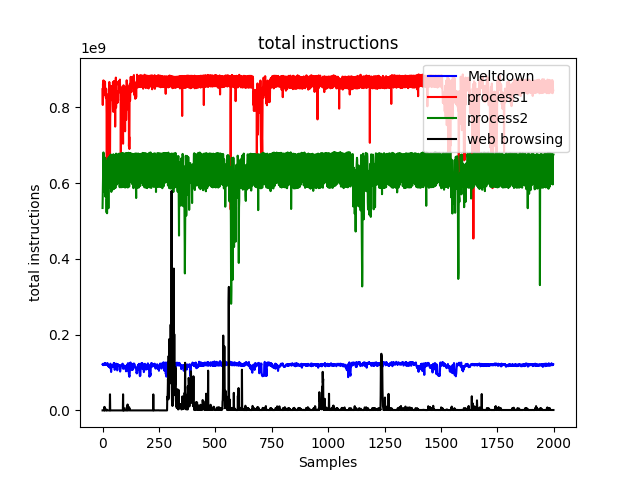}
\caption{Total instructions}
\label{fig7}
\vspace{-2ex}
\end{figure}

For Meltdown attack detection, we used both hardware and software events as features to machine learning models. We use Perf tool to monitor events related to Meltdown attack, because PAPI only provides hardware events and Perf support both hardware and software events. As mentioned in section \ref{metldown}, Meltdown is a two-step attack. In first step, Meltdown bypasses the memory isolation by executing unprivileged instruction out-of-order. In second step, it performs cache-based side channel attack to observe foot prints of accessed data from cache. Meltdown leaks secret information from user space and Kernel space by executing out-of-order memory lookups for desired reference memory. These unprivileged memory lookups generate segmentation faults or invalid page faults. Meltdown use exception handing to control the flow of code from being transferred to operating system, which saves the application termination. However, this page fault signal is still reported to operating system. We use perf to monitor this event related to page faults. We experimentally show that the Meltdown attack produce significantly higher page faults than other benign processes as shown in figure \ref{fig7}. Therefore, we select software event related to page fault for step one of Meltdown attack. Table \ref{tablemltdown} shows the selected hardware and software performance counters for Meltdown attack. 

\begin{figure}[htbp]
\centering
\includegraphics[width=0.40\textwidth]{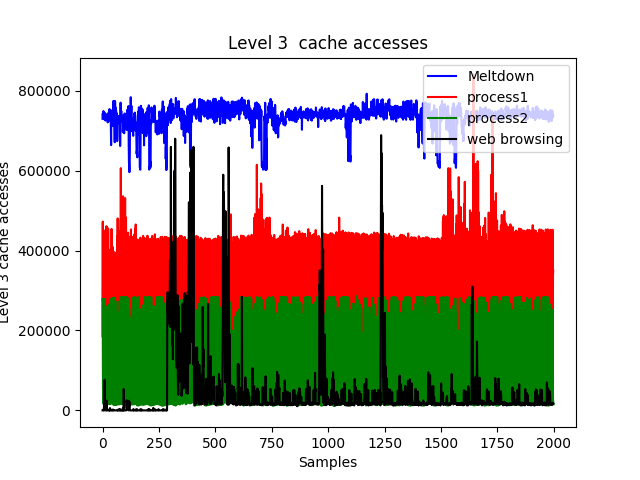}
\caption{Level 3 cache accesses}
\label{fig8}
\vspace{-2ex}
\end{figure}

\begin{figure}[htbp]
\centering
\includegraphics[width=0.40\textwidth]{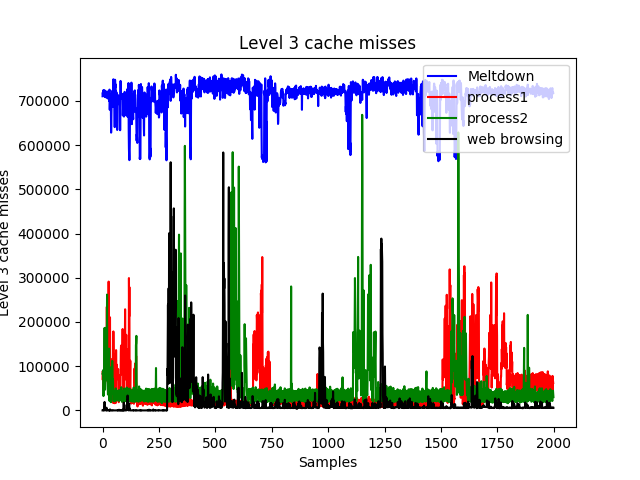}
\caption{Level 3 cache misses}
\label{fig9}
\vspace{-2ex}
\end{figure}

Like Spectre attack, in step two, Meltdown also executes cache-based side channel attack such as FLUSH+Reload to observe footprints of data that is accessed during out-of-order execution. We also select hardware events related to cache-based side channel attack to make detection mechanism for Meltdown more effective. Like Spectre attack, we select same events for second stage of Meltdown attack namely; L3 cache misses, L3 cache accesses and total number of instructions due to same reasons as discussed in section \ref{spectredetction}. Figure \ref{fig7} through \ref{fig9} show that Meltdown attack puts a very small workload, in the form of total number of executed instructions, on CPU but still generates higher cache misses, cache accesses. 

\vspace{-1ex}
\section{Experiments and Discussion}
\label{sec:experiments}
\vspace{-1ex}
For experimental verification and to check effectiveness of our proposed detection techniques, we perform three experimental case studies for the detection of Spectre variant 1, Spectre variant 2 and Meltdown attack. As mentioned in section \ref{rtd}, we use four machine learning model. To evaluate the performance of these machine learning models, we perform the experiments under realistic load conditions. We perform experiment under No Load (NL), Average Load (AL) and Full Load (FL) conditions by using SPEC integer benchmark \cite{news6}. SPEC integer benchmark provides system workload on branch predictor and memory intensive computations. We run the SPEC integer benchmark in the background as both Meltdown and Spectre attacks would also be affecting caches, branch predictors and out-of-order execution. This benchmark would create realistic system load conditions for the evaluation of our proposed detection techniques. In this section, we present the experimental results for two Spectre variants and Meltdown attack detection under variable workload conditions on CPU. We consider three factors to evaluate the performance of our proposed detection techniques, such detection accuracy, detection speed and performance overhead. 

We perform experiments with sampling rate of 100ms to extract performance counter values. But PAPI also support more fine-grained resolution. PAPI API can be used to get sampling rate of up to 3 microseconds. Increasing sampling rate increase the detection speed. However, it will increase the detection overhead. Our detection mechanism offers fine and coarse-grain profiling modes, which provide a trade-off between detection speed and impact on performance. Fine-grain mode collects samples at a very high frequency it is able to detect Meltdown and Spectre attacks relatively early with increased performance overhead. Whereas, coarse-grain mode takes samples at a low frequency, which takes relatively longer to detect the attacks, but it offers minimal performance overhead. Our detection technique can detect attacks in both modes. For experimental evaluation, we use 100ms sampling interval and show that it causes detection overhead less than 2\%. 

\vspace{-1ex}
\subsection{Detecting Spectre variant 1}
\label{subsec:FnRdetection}
Table \ref{tab:spectre1} shows  shows our experimental results for the detection of Spectre attack (Variant 1). All ML models provide consistently good detection accuracy under NL condition, i.e., up to 99.97 for LR, 99.25\% for SVM, 99.93\% for LDA and 99.93\% for CNN. Under AL condition, detection accuracy of LDA and CNN does not vary, but the accuracy of LR and SVM decreased to 98.40\% and 97.29\%, respectively, which still remains very high. Under FL condition, accuracy of all ML models drops by 2−3\% as compared to NL condition, except for the LDA. Jonas et al. in [25] have measured detection accuracy of Spectre attack under NL condition only. Our detection technique shows a very good detection accuracy for Spectre Variant 1 under NL, AL and FL conditions. Reduction in detection accuracy under AL and FL conditions are due to full utilization of shared resources (caches \& branch predictors) by benchmarks, benign processes and Spectre attacks. Detection overhead is measured to be between 1.67\%, 1.83\%, 1.89\%  3.51\% for LDA, LR, SVM and CNN respectively under all load conditions. As expected the detection overhead is highest for CNN due to its high computational requirement. 

\begin{table}[htbp]
\renewcommand{\thetable}{\arabic{table}}  
\vspace{1ex}
  \centering
    \caption{\textbf{Detection results using LDA, LR, SVM \& CNN models for Spectre Attack (Variant 1).}}
       \begin{tabular}{|r|r|r|r|r|r|r|}
    \Xhline{2\arrayrulewidth}
    \textbf{Model} & \textbf{Loads} & \textbf{Accuracy} &  \textbf{Speed} & \textbf{FP} & \textbf{FN} & \textbf{Overhead} \\ 
    &  & (\%) & ms & (\%) & (\%) & (\%) \\
    \Xhline{2\arrayrulewidth}
  \multirow{3}*{LDA} & NL &	99.93 &	100 & 0.07  &	0   & \\  \cline{2-6}
                     & AL &	99.06 &	100 & 0.57 &	0.37 & 1.67 \\ \cline{2-6}
                     & FL &	98.03 &	100 & 1.18  &	0.79   &  \\   \hline 
  \multirow{3}*{LR} & NL &	99.97 &	100 & 0.03  &	0   & \\  \cline{2-6}
                     & AL &	98.40 &	100 & 1.27 &	0.33 & 1.83 \\ \cline{2-6}
                     & FL &	97.36 &	100 & 1.98    &	0.66    &  \\   \hline
  \multirow{3}*{SVM} & NL & 99.25 &	100 & 0.69  &	0.06   & \\  \cline{2-6}
                     & AL &	97.29 &	100 & 2.02  &	0.69 & 1.89 \\ \cline{2-6}
                     & FL &	95.87 &	100 & 2.87    &	1.26    &  \\   \hline
  \multirow{3}*{CNN} & NL &	99.80 &	100 & 0.17  &	0.03  & \\  \cline{2-6}
                     & AL &	98.13 &	100 & 0.57 &	0.29 & 3.51 \\ \cline{2-6}
                     & FL &	97.43 &	100 & 1.56   &	1.01    &  \\   \hline 
       
  \end{tabular}
 \label{tab:spectre1}
\end{table}

\vspace{-1ex}
\subsection{Detecting Spectre variant 2}
\label{subsec:FnRdetection}
As mentioned in section \ref{spectre}, All variants of Spectre attacks generate significantly large amount of branch mis-predictions as compared to other processes having same or a smaller number of branch instructions. Therefore, to verify the effectiveness of our detection method, we also perform experiment with Spectre variant 2. Table \ref{tab:spectre2} shows  experimental results on the detection of Spectre Variant 2. Detection speed and overhead remain the same for both variants as the sampling granularity has been kept the same. Detection accuracy also remains almost the same, with the exception of a minor variation for CNN under FL condition. All ML models provided consistently good detection accuracy under NL condition, i.e., up to 99.98\% for LR, 99.38\% for SVM, 99.92\% for LDA and 99.37\% for CNN. Under AL and FL conditions, CNN performed better than all other ML models. All ML models report reasonably high detection accuracy for both variants of Spectre attack. The reported FPs are up to 2.87\% while FNs are up to 1.26\% only for both variants.

\begin{table}[htbp]
\renewcommand{\thetable}{\arabic{table}}  
\vspace{1ex}
  \centering
    \caption{\textbf{Detection results using LDA, LR, SVM \& CNN
models for Spectre Attack (Variant 2)}}
       \begin{tabular}{|r|r|r|r|r|r|r|}
    \Xhline{2\arrayrulewidth}
    \textbf{Model} & \textbf{Loads} & \textbf{Accuracy} &  \textbf{Speed} & \textbf{FP} & \textbf{FN} & \textbf{Overhead} \\ 
    &  & (\%) & ms & (\%) & (\%) & (\%) \\
    \Xhline{2\arrayrulewidth}
 \multirow{3}*{LDA}  & NL &	99.92 &	100 & 0.08  &	0  & \\  \cline{2-6}
                     & AL &	98.13 &	100 & 1.14 &	0.73 & 1.89 \\ \cline{2-6}
                     & FL &	98.23 &	100 & 1    &	0.77    &  \\   \hline     
  \multirow{3}*{LR}  & NL &	99.98 &	100 & 0.02  &	0  & \\  \cline{2-6}
                     & AL &	98.69 &	100 & 1.04 &	0.27 & 1.67 \\ \cline{2-6}
                     & FL &	98.13 &	100 & 1.41    &	0.46    &  \\   \hline
  \multirow{3}*{SVM} & NL &	99.38 &	100 & 0.57  &	0.05   & \\  \cline{2-6}
                     & AL &	98.29 &	100 & 1.28 &	0.43 & 1.83 \\ \cline{2-6}
                     & FL &	96.67 &	100 & 2.30    &	1.03    &  \\   \hline
    
  \multirow{3}*{CNN} & NL &	99.43 &	100 & 0.48  &	0.09   & \\  \cline{2-6}
                     & AL &	99.17 &	100 & 0.55 &	0.28 & 3.51 \\ \cline{2-6}
                     & FL &	98.69 &	100 & 0.79    &	0.52    &  \\   \hline 
       
  \end{tabular}
 \label{tab:spectre2}
\end{table}

\vspace{-1ex}
\subsection{Detecting Meltdown}
\label{subsec:FnRdetection}

\begin{table}[htbp]
\renewcommand{\thetable}{\arabic{table}}  
\vspace{1ex}
  \centering
    \caption{\textbf{ Detection results using LDA, LR, SVM \& CNN
models for Meltdown Attack}}
       \begin{tabular}{|r|r|r|r|r|r|r|}
    \Xhline{2\arrayrulewidth}
    \textbf{Model} & \textbf{Loads} & \textbf{Accuracy} &  \textbf{Speed} & \textbf{FP} & \textbf{FN} & \textbf{Overhead} \\ 
    &  & (\%) & ms & (\%) & (\%) & (\%) \\
    \Xhline{2\arrayrulewidth}
  \multirow{3}*{LDA} & NL &	99.95 &	100 & 0.05  &	0   & \\  \cline{2-6}
                     & AL &	99.83 &	100 & 0.13  &	0.04 & 1.79 \\ \cline{2-6}
                     & FL &	98.27 &	100 & 1.24  &	0.49    &  \\   \hline 
  \multirow{3}*{LR} & NL &	99.35 &	100 & 0.65  &	0   & \\  \cline{2-6}
                     & AL &	97.39 &	100 & 1.98  &	0.63 & 1.83 \\ \cline{2-6}
                     & FL &	94.67 &	100 & 3.43  &	1.90    &  \\   \hline
  \multirow{3}*{SVM} & NL &	99.97 &	100 & 0.03  &	0   & \\  \cline{2-6}
                     & AL &	99.17 &	100 & 0.67  &	0.16 & 1.91 \\ \cline{2-6}
                     & FL &	98.24 &	100 & 1.39  &	0.37    &  \\   \hline
\multirow{3}*{CNN}   & NL &	99.43 &	100 & 0.48  &	0.09   & \\  \cline{2-6}
                     & AL &	98.17 &	100 & 0.55  &	0.28 & 3.67 \\ \cline{2-6}
                     & FL &	98.69 &	100 & 0.79  &	0.52    &  \\   \hline 
       
  \end{tabular}
 \label{tab:meltdown1}
\end{table}

Table \ref{tab:meltdown1} shows experimental results of Meltdown detection using four machine learning models. Detection of Meltdown attack has not been reported in the state-ofthe-art as yet. Thus, to the best of our knowledge, these are the very first results illustrating how Meltdown attack can be detected at runtime using hardware and software performance events coupled with
machine learning models. While creating exceptions due to out-oforder execution, Meltdown attack generates distinctively higher page
faults. These page faults can be monitored through SPCs, whereas
HPCs can help monitoring other variations at the cache and branch
predictor levels. Results in Table II show that all four ML models
perform consistently well under variable load conditions. For instance,
LDA offers up to 99.95\%, LR offers 99.35\%, SVM offers 99.97\%,
and CNN offers 99.43\% detection accuracy for NL condition. Under
AL condition, detection accuracy of LDA, SVM and CNN remains
consistent, but the accuracy of LR model decreases to 97.39\%. Similar
behavior can be observed for FL conditions, where the accuracy of LR
is reduced by 4.68\% and that of SVM is reduced by 1.5\% as compared
to NL condition. Both LDA and CNN show consistent results under
variable system load conditions. The performance overhead of our
detection mechanism is only 2.11\% in the worst case for Meltdown
attack. Detection overhead \& speed for various ML models remains
the same due to fixed sampling granularity of 100ms. Due to high
detection accuracy, the reported error is very low, i.e., FPs are up to
3.4\% while FNs are up to 1.9\% only. These FPs and FNs are due to
non-deterministic nature of performance counters.

\section{Discussion on False positives and False negatives}
As shown in table \ref{tab:spectre1} through \ref{tab:meltdown1} , experimental results also report False positives and False negative for each ML model. False positives are higher for each ML model than False negatives. Higher False positive indicate that the ML models detect malicious processes with more accuracy than benign processes. These False positives and False negatives are due to non-deterministic nature of performance counters and Non-determinism varies values of hardware performance counters from 1-3\% [14]. Tools used to measure these performance counters are becoming more accurate with every latest version like PAPI or Perfmon. However, False positives and False negatives are outliers in data that never happens consecutively. In case of False negatives, if our detection service can’t be able to detect a Spectre/Meltdown with first time counters monitored values, it will be able to detect in next 100ms. Because the sampling rate of selected is 100ms. Also, False negatives are very less. In case of False positive, the results could be more disturbing, because detection service wrongly predicts a benign process as a malicious process. False positives and False negatives are very less in number as compared to detection accuracy of more than 99\%. 

As mentioned in section \ref{dectetionscheme} , our detection service does not handle malicious processes after they have identified. But we can make detection service more feasible by combining it with a application that handles these malicious processes. To address these issues of False positives and False negatives, our detection service passes the PIDs of processes predicted as malicious to an application which decides what to do with these processes. One idea is that instead of killing the process, halt the process and let the user decide what to do with these processes. This would make sure that no processes get killed due to wrong prediction in case of False positives and False negatives. Although, it is the responsibility of a mitigation technique to handle results of detection service but combining mitigation and detection techniques can provide better detection results and also decrease False positives and False negatives.

\section{Conclusion}
\label{conclusion}
This paper presents novel run-time detection mechanism for Spectre and Meltdown attacks on Intel’s x86 architecture. We perform experiments with four ML models under realistic system load conditions. These ML models use data from hardware and software performance counters to find out malicious behavior of Spectre and Meltdown attacks. We presents experimental evaluation of our proposed detection technique for Spectre variant 1, Spectre variant 2 and Meltdown attack.To the best of our
knowledge, this is the first-ever detection technique for Meltdown. Our results show that both Spectre and Meltdown attacks significantly alter the system’s run-time behavior by generating page faults, instruction mispredictions and cache accesses \& misses. These variations can lead toward early detection of such attacks using machine learning techniques. Our results report detection of Spectre and Meltdown with greater than 99\% accuracy with minimum performance overhead. 
 
Although, we perform experiments with Spectre variant 1, Spectre variant 2 and Meltdown. However, we believe that our proposed detection techniques will be able to detect other variants of Spectre and all other attack which exploits branch predictors, out-of-order execution, and cache based side channel attacks.

\bibliographystyle{IEEEtran}
\bibliography{ref}

\end{document}